\documentclass[runningheads]{llncs}
\usepackage[T1]{fontenc}

\usepackage{subfig}
\usepackage{multirow}
\usepackage{algpseudocode}
\usepackage{enumitem}
\usepackage{graphicx}

\usepackage{algorithm}
\usepackage{url}

\usepackage[overload]{empheq}
\usepackage{amsmath,amssymb}
\begin{document}
\title{A Reinforcement Learning–Inspired Latent Yield-based Adaptive Algorithm Switching Mechanism}
\titlerunning{RL-Inspired Latent Yield-based Adaptive AS Mechanism}

\author{
Jayprakash S. Nair\inst{1}\orcidID{0009-0008-6340-7250}\thanks{Corresponding author}
\and
Jimson Mathew\inst{1}\orcidID{0000-0001-8247-9040}
\and
Shivashankar B. Nair\inst{2}\orcidID{0000-0002-5246-286X}
}

\authorrunning{J. S. Nair et al.}

\institute{
Indian Institute of Technology Patna, Bihar 801106, India\\
\email{jsnair.hi@gmail.com, jimson@iitp.ac.in}
\and
Indian Institute of Technology Guwahati, Assam 781039, India\\
\email{sbnair@iitg.ac.in}
}
\maketitle              

\begin{abstract}
Selecting the most suitable algorithm for a given problem instance remains a challenging task, particularly in online or dynamic environments where problem characteristics evolve over time. Relying solely on instantaneous performance metrics can result in a reactive and unstable behaviour, often leading to suboptimal algorithm switching. 
This paper introduces a computationally efficient approach for aggregating an algorithm’s performance across multiple problem instances that is fairly immune to erratic variations in instance features. Inspired by features inherent to Reinforcement Learning (RL), this technique encapsulates rewards and penalties into a \textit{latent yield} that, in turn, triggers exploitation and exploration, consequently resulting in adaptive algorithm switching. The proposed technique employs island models, inspired by Genetic Algorithms, to facilitate parallel exploration and performance exchanges among algorithm populations inhabiting local repertoires. Experimental evaluations on sorting algorithms and robotic obstacle-avoidance tasks demonstrate the feasibility and effectiveness of the approach, highlighting its potential in domains where adaptive algorithm selection is critical.

\keywords{Adaptive Algorithm Selection \and Island Models \and Latent Yield \and Intrinsic and Extrinsic Exploration \and Reinforcement Learning \and Genetic Algorithms \and Multi-Agent Reinforcement Learning \and Multi-Agent Systems\and Robotics.}
\end{abstract}
\section{Introduction}
In the realm of AI, Algorithm Selection (AS) is a hot area that endeavours to choose the near-best algorithm given a set of candidate algorithms that can solve a problem instance \cite{kotthoff2016algorithm}. The problem of AS concerns identifying the most appropriate algorithm for solving a given problem instance or a class of problems. A typical AS comprises of a repertoire of algorithms, a problem instance space, a performance measure space and a feature space, all of which play a vital role in facilitating the switch to another algorithm. Despite the abundance of algorithms for a given computational task, their performance can vary drastically across instances due to factors such as input structure, resource constraints, and problem scale. Selecting an unsuitable algorithm may lead to wasted computation and suboptimal results, whereas a well-chosen algorithm can deliver significant performance improvements in efficiency, accuracy, and robustness. The need for AS arises out of the fact that no single algorithm performs best across all problem instances (No Free Lunch Theorem, NFLT) \cite{wolpert2002no}. This is especially true in cases where the environment is dynamic, where sticking to a pre-decided algorithm may not be worthwhile. Influenced by developments in machine learning, meta-optimization, and adaptive systems, the field of AS has evolved significantly and, as such, forms a cornerstone of automated machine learning (AutoML), evolutionary computation, and intelligent control systems, enabling computational systems to learn and adapt their behaviours based on empirical evidence.

AS can be performed either in an offline or online manner. Most of the work is concentrated on the former method. However, if the mapping of the algorithms does not capture the nature of all types of instances, then poor selections are bound to degrade performance. Offline algorithm selection methods neglect the information regarding the performance of the selected algorithm on new online instances \cite{degroote2017online}.
AS has been applied in various domains, including Graph Neural Networks (GNNs) \cite{yuan2024enhancing}, data-driven models \cite{rook2025efficient}, and multi-armed bandit algorithms \cite{tornede2022machine}.
It can be used to train an ANN to select the best algorithm. Traditional algorithm selection methods assume static problem conditions, but many real-world systems, such as cloud computing, robotics, and streaming data processing, require online adaptive selection of algorithms. In such systems, the algorithm choice needs to be revised continuously based on feedback from the environment. Reinforcement Learning (RL) and Bayesian optimization have proven useful for such adaptive strategies \cite{lindauer2015autofolio}.
Evolutionary computation and Population-Based Training (PBT) methods evolve algorithm portfolios, share performance information, and co-adapt algorithmic behaviour to dynamic conditions \cite{kerschke2019automated}. These hybrid frameworks become especially relevant when it comes to optimization and autonomous decision-making systems.

In this paper, we propose a novel and computationally simple method for online algorithm switching that constantly monitors the performance of the algorithm, accumulates and converts it into a latent yield. This latent yield, which constitutes a memory of the aggregate of the past performances of the algorithm, is used to make a decision on whether or not to perform a switch. 
While this latent yield is synonymous to a combination of rewards and penalties, just as in a typical RL algorithm, the proposed method also inherently performs exploitation and exploration. 
While this latent yield is synonymous to a combination of rewards and penalties, the proposed method also inherently performs exploitation and exploration, making it akin to RL. In addition, the method makes use of parallelism by conforming to the island models used in Genetic Algorithms (GA). This makes it scalable, both in terms of parallelism and an increase in the number of algorithms within the respective algorithm repertoires. The significant contributions of the work described in this paper include –

\begin{enumerate}[nosep]
    \item 	A cumulative latent yield that remembers the overall performance of an algorithm
    \item Delayed switching, thereby avoiding reflex action
    \item Parallelism in AS using island models
    \item Scalability in terms of an increase in both the number of islands and new algorithms
\end{enumerate}

Subsequent sections describe the main methodology, system dynamics, exploration and exploitation mechanisms, followed by results when the AS was applied to a closed and open world problem. 

\section{Related Work}

The algorithm selection problem was first formalized by Rice \cite{rice1976algorithm}, who proposed mapping problem instance features to the most suitable algorithm. Early approaches focused on \emph{offline selection}, where algorithms were chosen once before execution based on static features or empirical benchmarking. With the advent of machine learning–driven selection, systems such as Satzilla \cite{xu2008satzilla} and ASlib \cite{bischl2016aslib} leveraged supervised learning to predict the best algorithm using feature-based meta-models. However, these methods operate in \emph{static or batch settings} and are unsuitable for online or continuously evolving environments, where problem distributions shift over time. Recent work has therefore explored adaptive and online AS, with RL and bandit-based methods enabling dynamic switching among algorithms during execution \cite{lindauer2018algorithm}. While effective, these approaches often rely on instantaneous reward feedback, which can cause instability or premature switching when performance fluctuates rapidly.

Island model architectures, originally developed in genetic algorithms \cite{whitley1999island}, have also been used to promote diversity in optimization by enabling parallel exploration with periodic migration. Similarly, decentralized and distributed frameworks inspired by biological systems have been proposed for online algorithm selection in networked environments. Semwal \cite{semwal2020decentralized} describes a decentralized Cyber–Physical System (dCPS) mechanism that evolves mappings between problems and solutions across a network of heterogeneous nodes. Their system continuously adapts to a stream of varying problems, shares evolved solutions among nodes, and demonstrates real-world applicability in both sorting tasks and robotic path-following. This work highlights the benefits of distributed knowledge sharing and continuous learning for selecting optimal solutions without relying on central coordination.

This paper situates itself at the intersection of these paradigms — proposing a latent yield–based cumulative estimator for algorithm performance, combined with an island-based exploratory–exploitative architecture, designed to prevent knee-jerk switching.

\section{Methodology}
\vspace{-0.25cm}
In computational science, optimization, and multi-agent systems, island models (also called distributed models or archipelagos) \cite{whitley1997island} refer to a way of structuring a population of agents in semi-independent groups called islands. A significant challenge in many problems is determining the optimal times to seek improved solutions within the island (exploitation) or from other islands (exploration). The proposed mechanism is designed to address this issue through a multi-agent framework organized as an archipelago, shown in Fig. 1. As can be seen, multiple islands exchange information via a Central Interface Agent (CIA). Each island has its own dedicated agent that can run an algorithm to cater to input task instances. The island also has a repertoire of algorithms and a Yielory that holds a latent yield, which is crucial to AS. We describe this in more detail below.

\begin{figure}[ht]
    \centering
    \includegraphics[width=0.5\linewidth, height=4cm]{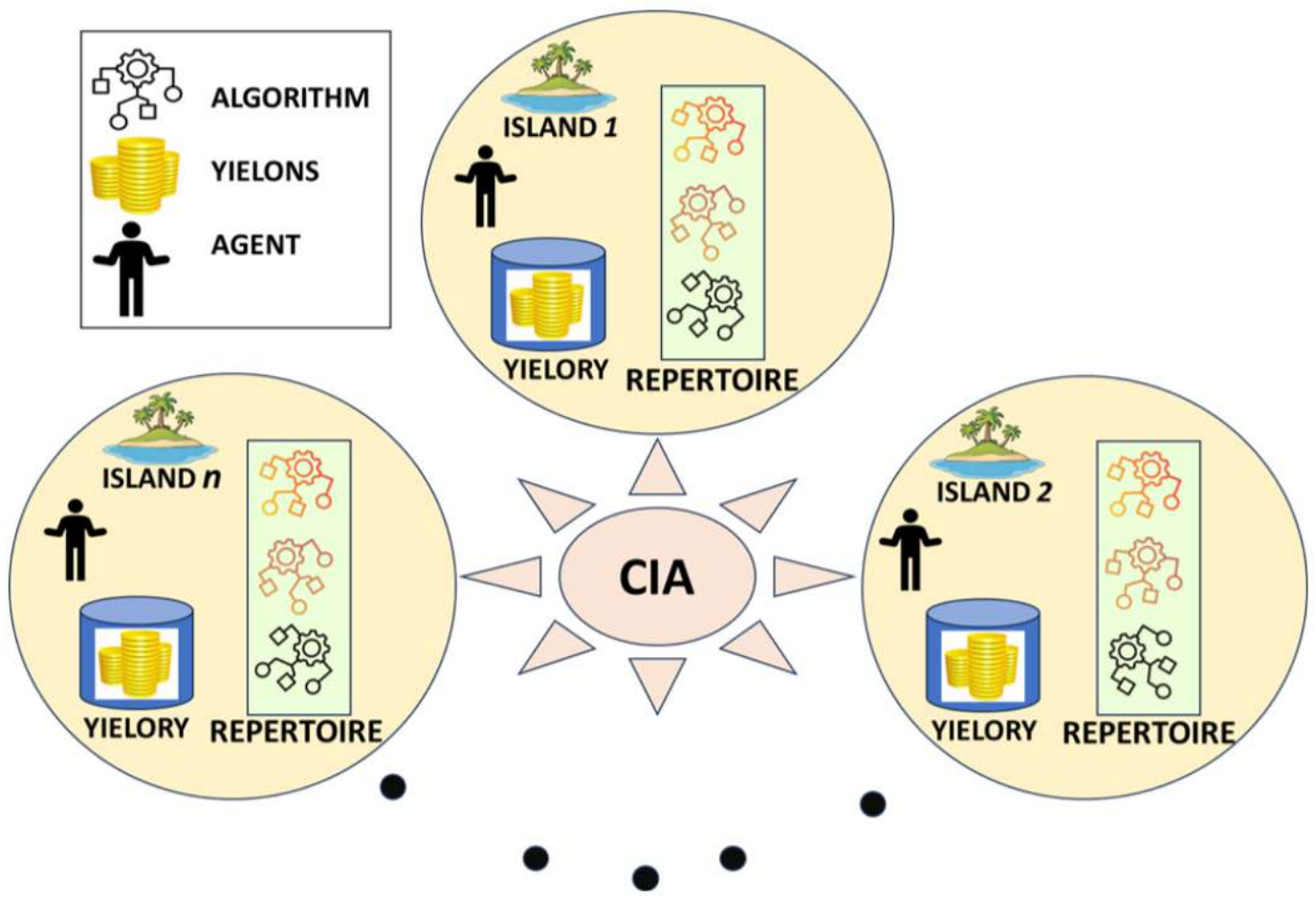}
    \caption{System architecture for the proposed AS Mechanism}
\end{figure}
\vspace{-4mm}

\vspace{0.1cm}
Let
${I} = \{ I_{1}, I_{2}, I_{3}, \ldots, I_{n} \}$ represents a finite set of $n$ islands, each of which could be considered as a separate computing device capable of hosting an island and its constituents. These islands communicate via a CIA situated in either of these islands or a separate computing device. The CIA does not actively impose decision rules on the agents but primarily observes and keeps track of the best-performing algorithm across the islands based on its performance. Each island $I_{i} \in \mathcal{I}$ is endowed with an agent $\psi_{i}$ capable of executing the concerned task for which the system has been designed. In addition, every island has a  \textit{repertoire} of algorithms, $\mathrm{Rep}_{i} = \{ A_{i1}, A_{i2}, \ldots, A_{i m_i} \}, \quad m_i = |\mathrm{Rep}_{i}|,$
where $m_i$ denotes the number of algorithms associated with the agent $\psi_{{i}}$ in $I_{i}$. All algorithms within all repertoires are designed to accomplish the same task, $T$. Agents within all the islands, thus, strive to find the best algorithm for the same task, $T$. Further, the repertoires are not necessarily disjoint. For distinct agents $\psi_{i}, \psi_{j} \in {\Psi}$, $i \neq j$, it is not necessary that,
${Rep}_{i} \cap \mathrm{Rep}_{j} \neq \phi$ indicates that repertoires of different islands could have some algorithms in common.

\subsection{Agent-side Actions}
Initially, an agent, $\psi_{i}$, in an island, $I_i$, randomly selects an algorithm from its repertoire, ${A_{\psi_i} \in Rep_{i}}$, available within its island.
The agent uses this algorithm to perform a series of task instances, \(x_j \in \mathcal{X}\), of the same task $T$, which it encounters. After performing each task  
instance $x_j$, the agent $\psi_{i}$, receives an extrinsic reward or score, which we term as \emph{credit}. This is proportional to the computational resources $R$, and time $\tau$, required to perform the task instance using the selected algorithm. Since different agents may execute different algorithms and task instances can also vary in their intrinsic parameters, the credit values could differ across agents. 
Further, since the algorithms within an agent's repertoire are heterogeneous, the credits earned by each of them on execution by the agent could lie in different scales. To allow comparison across these algorithms in the repertoires,
we define a normalized credit as -
\begin{equation}
C_{\psi_i, A_j}^{\mathrm{norm}}(x_j)
=
\frac{
    k / f(R_{A_j}, \tau_{A_j}, x_j)
}{
    \max_{x_k \in \mathcal{X}_{A_j}} \left( k / f(R_{A_j}, \tau_{A_j}, x_k) \right)
}
\times 100
\label{norm credit}
\end{equation}
where $f(C_{A_j}, \tau_{A_j}, x_j)$ is a task-specific cost function, and $k$ is a positive scaling constant, 
$\mathcal{X}_{A_j}$ denotes the set of all task instances executed by the agent, $\psi_i$, while using $A_j$, and $x_k$ refers to 
the instance of task for which the algorithm performed the best. This normalized credit provides a percentage measure of 
performance of the algorithm, which is task-specific and suitable for use in subsequent aggregation steps. It is utilized to compute a latent yield, described in the next section, similar to the one used by Jha \cite{jha2014stigmergy}. This yield has the potential to generate an innate drive within an agent, which in turn, assists it in determining whether or not to switch from the currently active algorithm to another. Algorithm switches can occur either internally (i.e., to an algorithm available in the repertoire within an island) or externally (i.e., to an algorithm in the repertoire of another island).
\subsection{Latent Yield}
We coin a term $Yielon$ ($\Upsilon$) to represent a unit for the latent yield produced by an agent after running an algorithm on an instance of a task. The dynamics of $\Upsilon$ mechanism is akin to wringing water from a towel soaked in water. The first wring with a torque $q$, can extract a high amount of water from the towel that acts as a water repository. To extract the same amount of water once again requires more exertion or a torque higher than $q$, unless the towel is \textit{charged}, by some means, with more water. The dynamics of the Yielon is based on a similar principle. If the algorithm run by an agent performs well (i.e., when its normalized credit gain is high), the system tends to recharge the Yielon repository or Yielory ($\Upsilon_{rep}$), proportionate to its current Yielon count, thereby increasing the number of Yielons. If the algorithm performance worsens, it tends to deplete the Yielons in the Yielory proportionate to its current content. The dynamics of this charging and depletion of Yielons is discussed in the following section.
\subsection{Algorithm Switching Dynamics}
Let $\Upsilon_{curr_i}$ be the Yielons accumulated by an agent $\psi_i$ while running an algorithm $A_i \in Rep_{i}$. After execution of every instance, $x_{j}$, of a task, the normalized credit is computed using the Eq.\ref{norm credit}. A set of recent normalized credit values,  $C_{\psi_i}^{\mathrm{norm}}(A_i, x_{j})$, is maintained using a sliding window of size, $w$.
The mean of the consecutive differences of these values within the window constitutes what is termed as the \textit{Squeezing factor}, $\sigma$, calculated as:
\vspace{-0.5cm}
\begin{equation}
\label{gamma}
\sigma = \frac{1}{w-1} \sum_{j=1}^{w-1} 
\big( C_{\psi_i}^{\mathrm{norm}}(x_{j+1}) - C_{\psi_i}^{\mathrm{norm}}(x_{j}) \big)
\end{equation}
\vspace{-0.5cm}

where:
\begin{itemize}[nosep]
    \item $w$ denotes the number of elements maintained in the normalized credit window of size ${W}$,
    \item $C_{\psi_i}^{\mathrm{norm}}(x_j)$ represents the normalized credit values obtained by agent, $\psi_i$, on performing the $j^{\text{th}}$ task instance $x_j$,
    \item the term $\big(C_{\psi_i}^{\mathrm{norm}}(x_{j+1}) - C_{\psi_i}^{\mathrm{norm}}(x_{j})\big)$ captures the local normalized credit gradient earned from consecutive instance performances.
\end{itemize}

This squeezing factor, $\sigma$, acts as the torque, and regulates the replenishment and depletion of Yielons stored within a \textit{Yielory}, whose contents in turn influence the decision to replace the current algorithm. The update of Yielons in the Yielory is performed based on Eqn: \ref{yielon approx} and \ref{yielon not approx}, given below. 

\begin{subequations}
\label{Phi:def}
  \begin{align}[left ={\Upsilon_{new} = \empheqlbrace}]
    &    \Upsilon_{curr} + \sigma * (\Upsilon_{curr} / \Upsilon_{max}), & \text{if $\sigma \not \approx 0$}   \label{yielon not approx}\\
    &\Upsilon_{curr} + p * (\Upsilon_{curr} / \Upsilon_{max}), & \text{if $\sigma \approx 0$ }  \label{yielon approx}
   \end{align}
  \end{subequations}

where, $\Upsilon_{curr}$ is the current Yielon count in the Yielory, $\Upsilon_{max}$ is its maximum capacity, and $p$ is a small positive value. A higher value of both $\sigma$ and $\Upsilon_{curr}$ would yield more Yielons to fill up the Yielory. From Eqn. \ref{Phi:def}, it is clear that when the current algorithm is faring badly, the extraction of more Yielons would require a substantially higher value of $\sigma$, much like the squeezing torque in a physical system. The Yielory thus, accumulates lesser Yielons. Conversely, if the Yielory was full, even a small amount of squeeze ($\sigma$) will cause the extraction of a large amount of Yielons to be added onto the Yielory. It may be observed that the first term on the R.H.S. of Eqn. \ref{Phi:def} indicates the current amount of Yielons in the Yielory, while the second form indicates the amount extracted from it. Since the Yielory grows or diminishes for every task instance performed, it serves as a sort of memory of how well the algorithm performed over a large period of time. From another perspective, this performance-dependent waxing and waning of the Yielons in the Yielory could be looked upon as a consolidated memory of the past performances learned through RL.
As mentioned earlier, $\sigma$, represents the mean rate of change in normalized credits over recent task
instances. A smaller value of $\sigma$ indicates that the agent’s performance is saturating (i.e., the learning process is converging), while a larger magnitude of the same implies an ongoing adaptation or instability in performance.
$\sigma$ is subsequently used to regulate the agent’s adaptive AS mechanism. If the
Yielory had a high number of Yielons, and the algorithm fared badly in the
last few instances; the system would still continue to use the currently active algorithm
(exploitation), since its Yielon count is rich enough to sustain this fall in performance. On
the contrary, if the Yielory count is sparse, which means the algorithm has been
performing badly for quite some iterations, the depletion rate of Yielons due to a
negative $\sigma$, is drastic. When this count reaches a lower threshold, it triggers an imminent switch to another algorithm within the local repertoire. The Yielon count thus avoids spontaneous or reflex-based switching of
algorithms. This gives the current algorithm more chances to thrive in a given environment. 
\subsection{Exploitation and Exploration}
In search, exploitation, and exploration 
need to be balanced based on past performances. The content of the Yielory,
in some way defines the performance of the system over a period of time. Yielons
in the Yielory grow and diminish based on the performance of the current algorithm. As long as the Yielory has a sufficient amount of Yielons to thrive, switching
to another algorithm is inhibited. This means the agent continues to remain satisfied and exploits the currently used algorithm even if its performance is a bit sporadic.
However, if the algorithm performs badly for a longer period of time, causing the
depletion Yielons beyond a certain lower minimum threshold, the agent is forced to explore and hence switch
to a new algorithm in the repertoire of its own island (\textit{intrinsic exploration}) or that from another (\textit{extrinsic exploration}). 
On the contrary, there could be other conditions when the Yielory gets filled to
its maximum and saturates to $\Upsilon_{max}$ due to continuous good performance exhibited by the algorithm, indicating a possible local optimum. The squeezing factor $\sigma$, computed using the mean gradient method as presented in Eqn. \ref{gamma}, determines whether an agent $\psi_i$ should exploit its currently active algorithm $A_{\psi_i}$ or explore a new one. The conditions for deciding this are given below:

\begin{enumerate}[nosep]
    \item When $\sigma \approx 0$, it signifies that the normalized gratuities within the window, $W$, have saturated. To decide whether to exploit or explore, the difference: 
        
        \begin{equation}
        \label{delta}
            \Delta = {C_{\psi_i}}_j^{\mathrm{norm}}({A_{\psi_i}}_j, x_j) - C_{min}^{\mathrm{norm}}
        \end{equation}
    is computed. Here, ${C_{\psi_i}}_j^{\mathrm{norm}}({A_{\psi_i}}_j, x_j)$ is the normalized credit of the current instance  $x_j$ and $C_{min}^{\mathrm{norm}}$ is a predefined minimum desirable normalized credit threshold.

    \begin{enumerate}[nosep]
        \item If $\Delta \ge 0$ , the agent continues to exploit the current algorithm. In this case, the Yielon count is updated using Eqn. \ref{yielon approx}, thereby increasing it marginally. 
        \item If $\Delta < 0$, the agent switches to exploration by querying the CIA for the best active algorithm across all islands, $A_{best}$, and its corresponding Yielon count, $\Upsilon_{best}$.

If either
\begin{equation}
{A}_{best} = {A}_{\psi_i} \quad \vee \quad \left| \Upsilon_{best} - \Upsilon_{curr} \right| < \epsilon    
\label{condition intrinsic}
\end{equation}

for some small threshold \( \epsilon > 0 \), then the agent opts for \textit{intrinsic exploration} by randomly selecting a new algorithm from its repertoire, $Rep_i$ as below:
\begin{equation}
\label{randomize algo}    
A_{new} \sim \text{Random}({Rep_i} \setminus \{{A}_{{\psi_i}}\})
\end{equation}

Else, it undertakes \textit{extrinsic exploration} by switching to the best algorithm, $A_{best}$.
    \end{enumerate}
    \item When the gradient of the normalized gratuities in the window is approximately non-zero, ($\sigma$ $\not\approx$ 0) the agent continues to \textit{exploit} the current algorithm. Thus, when the Yielon count increases ($\sigma$ > 0) or decreases ($\sigma$ < 0) and remains between the $\Upsilon_{max}$ and $\Upsilon_{min}$, the agent sticks to exploiting the current algorithm as per Eqn. \ref{yielon not approx}.

    \item When $\Upsilon_{new}$ < $\Upsilon_{min}$, the agent switches to another algorithm within its repertoire exhibiting \textit{intrinsic exploration}.

\end{enumerate}

The exploitation strategies allow the agent time to improve the performance and avoid any impulsive change of algorithm. Algorithm \ref{algo} shows the algorithm for the proposed algorithm switching mechanism. 

\vspace{-2mm}
\small
\begin{algorithm}[ht]
\caption{The proposed AS mechanism}
\label{algo}
\begin{algorithmic}[1]
\Require $\mathrm{Rep}_{i} = \{ A_{1}, A_{2}, \ldots, A_{m_i} \}$
\State \textbf{Initialize:} $\Upsilon_{max}$, $\Upsilon_{min}$, $\Upsilon_{curr}$, $\Upsilon_{initial}$
\State Assign an algorithm $A_j \in \mathrm{Rep}_{\psi_i}$
\State $\Upsilon_{curr} = \Upsilon_{initial}$
\For{each instance, $x_j \in X$ of task $T$}
    \State Execute algorithm $A_j$ on instance $x_j$
    \State Measure computational effort $R$ and execution time $\tau$
    \State Compute Normalized credit, $C_{\psi_i, A_j}^{\mathrm{norm}}(x_j)$ (using Eqns. \ref{norm credit})  
    \State Compute $\sigma$ (using Eqn.~\ref{gamma})
        
    \If{$\sigma \not \approx 0$}
        \State Update $\Upsilon_{new}$ (using Eqn.~\ref{yielon not approx})
   
    \Else
        \State Compute $\Delta$ (using Eqn. \ref{delta})
        \If{$\Delta \geq 0 $}
            \State Update $\Upsilon_{new}$ by a small threshold $p$ (using Eqn. ~\ref{yielon not approx})
        \Else
            \State Query CIA for $A_{best}$, $\Upsilon_{A_{best}}$
            \State $\Upsilon_{curr} = \Upsilon_{initial}$
            \If {Condition satisfied (\ref{condition intrinsic})}
                \State Select $A_{new}$ using Intrinsic Exploration (Eqn. \ref{randomize algo})
            \Else 
                \State $A_{new}$ $\leftarrow$ $A_{best}$ (Extrinsic Exploration)
            \EndIf
        \EndIf
    \EndIf
    \If{$\Upsilon_{new} < \Upsilon_{min}$}
        \State  $\Upsilon_{curr} = \Upsilon_{initial}$
        \State Select $A_{new}$ using Intrinsic Exploration (Eqn. \ref{randomize algo})
    \EndIf
\EndFor
\end{algorithmic}
\end{algorithm}

\subsection{Galapagos Island}

If all agents across the islands were to behave similarly, the overall search process could stagnate around a local optimum, with all islands converging to identical or nearly equivalent algorithms in terms of performance. To prevent such a premature convergence, it is essential that at least one island hosts an agent that exhibits a deviation in exploratory behaviour which is distinct from the rest. This agent, residing on what we term the \textit{G-Island}, (G for Galapagos) actively promotes diversity across the islands, by evaluating comparatively unexplored or underutilized algorithms. 
When the agent within \textit{G-Island}, $I_G$, initiates exploration, the new algorithm, $A_{new}$, is selected based on the conditions outlined below.
\vspace{-10pt}
\begin{equation}
\begin{aligned}
(I = I_G)\land
\Big[
&\big(\Upsilon_{\mathrm{mean}} - \Upsilon_{I_G} > \delta\big)
\\
&\lor
\Big(
I_G = \operatorname*{arg\,max}_{I_i\in\mathcal{I}}\!\mathrm{Perf}(I_i)
\land
\exists\,I_j\neq I_G:\;A_j = A_G
\Big)
\Big]
\\[4pt]
&\Rightarrow\;
A_{\mathrm{new}}
=
\operatorname{Random}\!\Big(
Rep_G \setminus \{A_i \mid I_i \in \mathcal{I}\}
\Big)
\end{aligned}
\end{equation}

It may be noted that the first condition indicates that the Yielories of all islands have reached their maximum capacities. The second ensures that it selects an algorithm from within its internal repertoire, which is different from the currently active algorithms used across the islands.

\section{Experimental Setup}
Experiments were performed using two different domains and a randomly assigned, Island-2 as the G-Island, with the following parameter values: $\Upsilon_{max}$ = 100, $\Upsilon_{min}$ = 30, $\Upsilon_{initial}$=60, $W$=5, $p$ = 0.05, $\Delta = 80$, and $\epsilon=10$. 
\subsection{Finding the Best Sorting Algorithm}
To test the proposed method, we utilized three sorting algorithms - Quick-sort (QS), Insertion Sort (IS), and Counting Sort (CS) - distributed across the repertoires of three distinct islands. The repertoires on each island contained these three sorting algorithms. Each island had its own randomizer function, which generated different data types of length 20 and fed them to the active sorting algorithm. The randomizers generated three types of input data at random:
\begin{enumerate}[nosep]
    \item \textbf{Randomized Large (RandL)}: An array of random numbers having a large range ($0, 10^6$), ideally suited for Quick-sort.
    \item \textbf{Randomized Small (RandS)}: An array of random numbers with a small range (1 to 35), suited for Counting Sort.
    \item \textbf{Almost Sorted (AlmoS)}: An array of nearly sorted numbers, perfect for Insertion Sort.
\end{enumerate}

The randomizers ensured that each active algorithm received different combinations of these input types. This allowed for a comprehensive evaluation of how each sorting algorithm performs with varying types of data and conditions, enabling the testing of efficiency and suitability of each algorithm for different types of input data or sorting instances. After each array was sorted, the associated credits were calculated using the following equation:
\vspace{-0.3cm}
\begin{equation}
C_{\psi_i, A_j}(x_j) = \text{round} \left( \frac{1}{\omega + \alpha + 1} \right)
\end{equation}
where \( \omega \) is the number of instructions executed, and \( \alpha \) represents the number of memory accesses. These credits were used to compute the normalized credits and subsequently the Yielon count, which in turn aids in switching the algorithms. For comparison, baseline experiments were conducted without using the Yielory and G-Island concept. The agents in the islands were initialized with a different algorithm and used a greedy AS strategy. After every episode, this greedy strategy explored the current best algorithm in the archipelago and made the agent switch to it.

\subsection{Learning to Avoid Obstacles} This experiment was carried out in the robotics domain to validate the working of the proposed method in near real-world scenarios. Here, the agents were substituted by identical robots inhabiting a simulated setup using Webots \cite{Webots04}. Four instantiations of Webots, each containing a robot, a Yielory, and a repertoire, formed the four islands. Each robot inhabited a different environment
with a variable number of static obstacles positioned at random locations. A snapshot of such an environment is shown in Fig. 2. The main
objective was to arrive at the right algorithm to avoid the obstacles and sustain
within the given environment. We have used three learning algorithms - Q-Learning (QL) \cite{watkins1992q}, SARSA \cite{rummery1994line}, and Double-Q-Learning (DQL) \cite{NIPS2010_091d584f}. Further, by conferring two different values of learning rate ($\alpha = 0.1$ and $\alpha = 0.7$) to each algorithm, we created a population of six algorithms in the repertoires. The robots, together with the RL algorithms, constituted a Multi-Agent Reinforcement Learning (MARL) system.
If the robot (agent) during its motion does not encounter an obstacle for a certain period of time (5 seconds), then the active algorithm used by it was conferred a Normalized credit of 10 units. 
In addition, the charge on the battery of each robot at start was kept at 100\% (maximum charge), which, on every action or movement made by the robot, decreased by 2.5\% of the maximum. When the charge depletes to zero, an episode that constitutes an instance of learning the task was assumed to have been completed. At the end of each such episode, the Yielory is updated and a new episode commences with the battery once again charged to the maximum.

\begin{figure}[ht]
    \centering
    \includegraphics[width=0.5\linewidth]{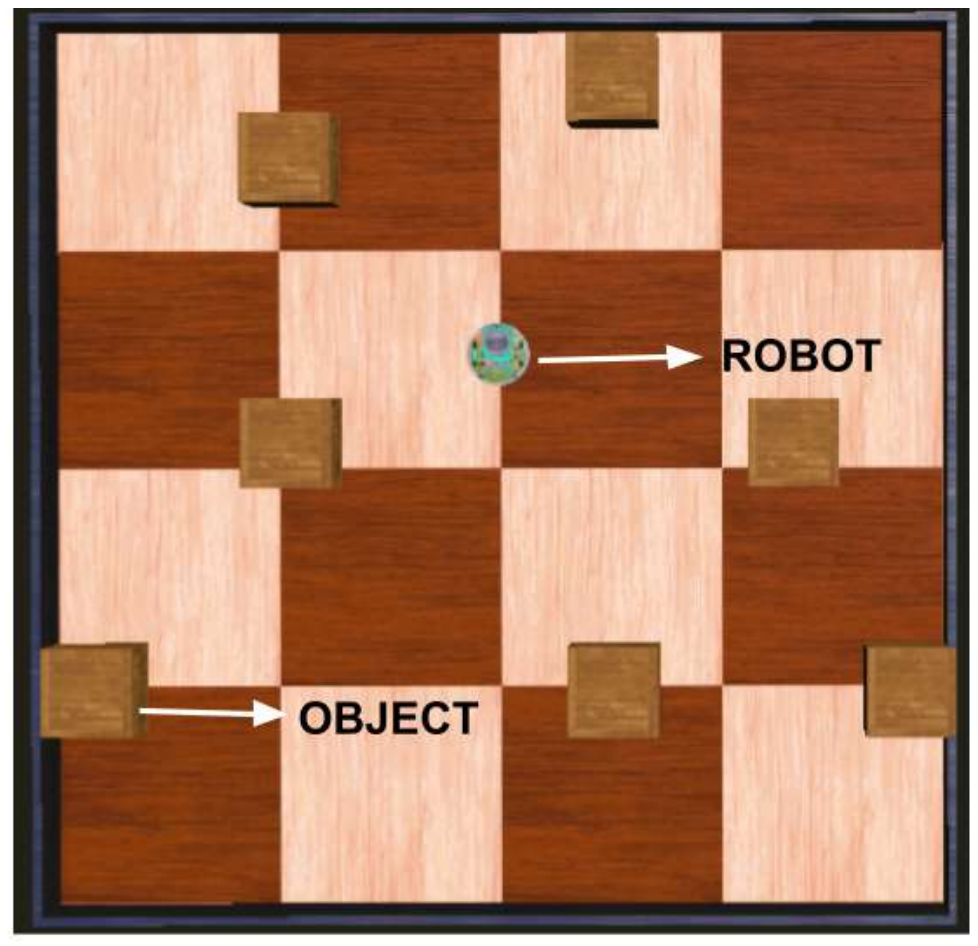}
    \caption{Snapshot of the environment of one of the Webots \cite{Webots04} instantiation arena featuring a robot amongst multiple static obstacles at different positions. 
    }
\end{figure}
\vspace{-4mm}

\section{Results}
In this section, we discuss the results based on the types of experiments conducted.

\vspace{-0.5cm}
\subsection{Sorting} Figs. 3a - 3c, depict the baseline graphs of variations in normalized credits versus episodes, across the three islands when a greedy approach coupled without and with a G-Island was used. 
The vertical black dashed lines indicate a switch to a new algorithm, mentioned alongside.
The vertical colour bands indicate the type of data input to the sorting algorithm during the episodes. In the graph in Fig. 3a, the agent in $I_1$ started off using CS. Since this algorithm is best suited for data of type RandS, it yielded a high value of normalized credit. $I_2$ started with IS, which is not compatible to RandL. It thus switched to CS, the best performing algorithm across all islands at that episode (Episode-2). A similar pattern can be observed at $I_3$, which also switched to CS (Episode-2). Eventually, all islands switched to CS and stuck to it in spite of varying data types. The islands thus quickly stabilized to the same algorithm, curbing exploration of potentially better algorithms, thereby resulting in a local optimum. 
In Figs. 3d - 3f, the G-Island switched frequently causing the other islands to copy its algorithm every time it was a better one. Subsequently, this caused other islands to also switch to the better one, at about the same rate as the G-Island. Thus, the introduction of the G-Island did not really do much to improve the search for the best algorithm or the number of switches. 
Fig. 4a - 4c depict the variations in normalized credits and Yielon count with episodes. It can be seen in Fig. 4a that even though there were erratic changes in data types in $I_1$, the presence of a substantial amount of Yielons generated an inertia that delayed switching to a new algorithm from the initial CS algorithm. It later switched to IS at the 100th episode when the Yielon counted dipped to its lowest, retained this algorithm since it was compatible to the data type (AlmoS) till the 200th episode when the data type changed to RandL and then switched many times to find a better performing algorithm. Similar behaviours can be observed in the other two islands (Figs. 4b and 4c) where the Yielon count controls the switching. 
Figs. 4d- 4f depict similar graphs when the G-Island was introduced. It can be clearly seen that in the G-island the number switches are far more than the others indicating that it has struggled to keep exploring more often than the other two islands.

\begin{figure}[t]
\centering

\textbf{No Yielory and No G-Island}\\[4pt]

\makebox[\textwidth][c]{%
\subfloat[$I_1$ (Start Algo.: CS)]{
  \includegraphics[width=0.32\textwidth]{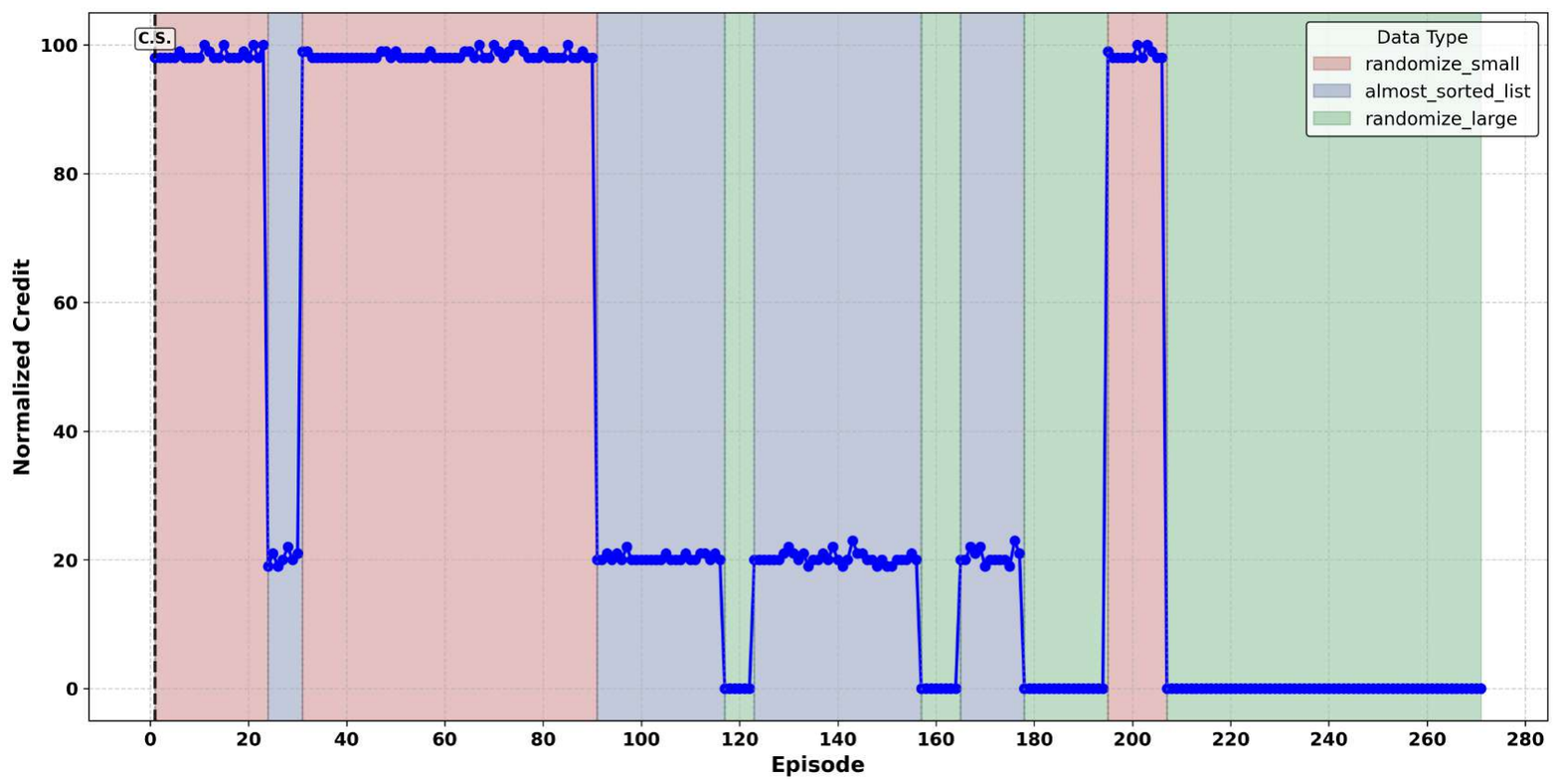}
}\hfill
\subfloat[$I_2$ (Start Algo.: IS)]{
  \includegraphics[width=0.32\textwidth]{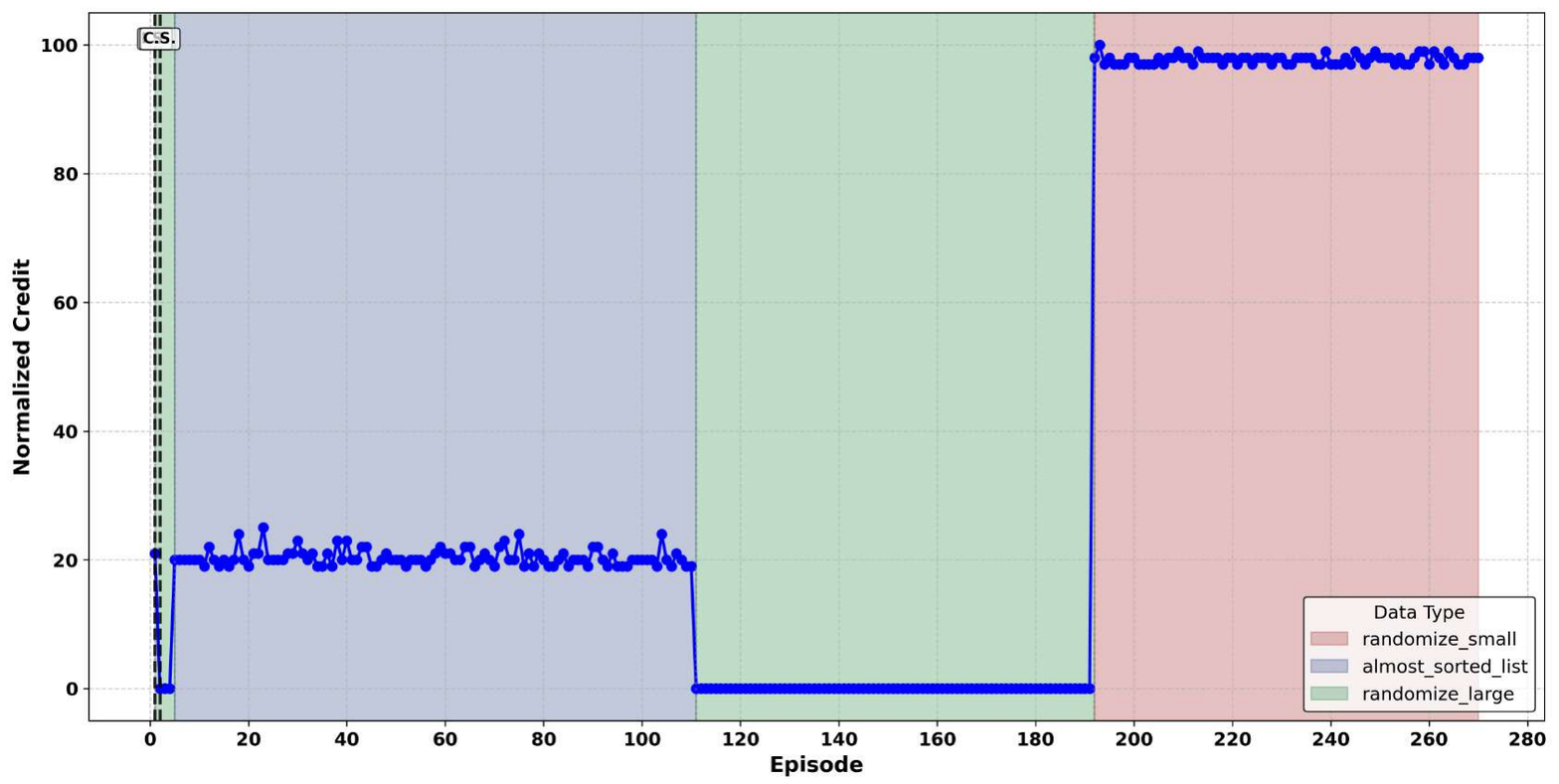}
}\hfill
\subfloat[$I_3$ (Start Algo.: QS)]{
  \includegraphics[width=0.32\textwidth]{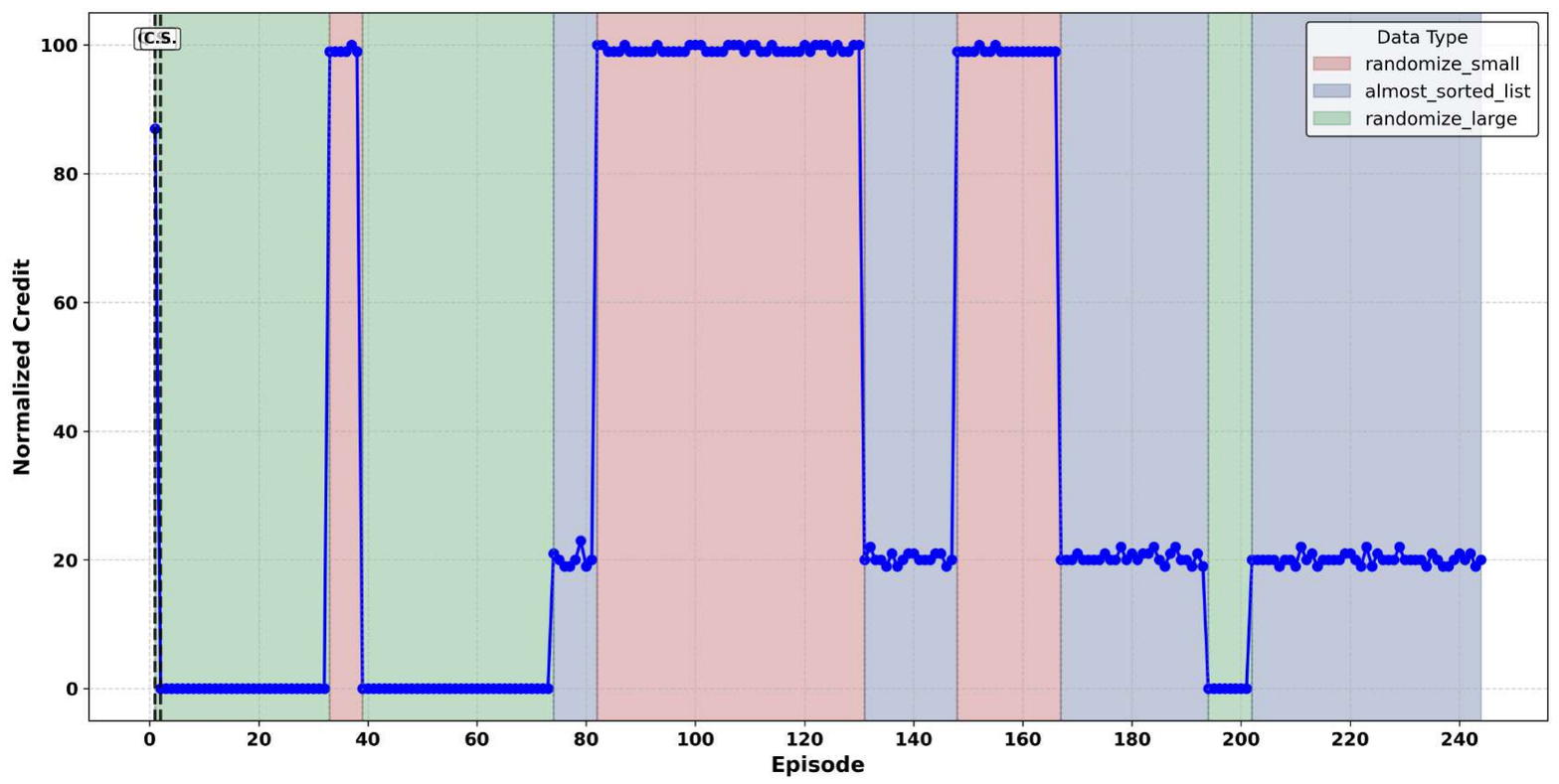}
}
}

\vspace{1em}

\textbf{No Yielory and With G-Island}\\[4pt]

\makebox[\textwidth][c]{%
\subfloat[$I_1$ (Start Algo.: CS)]{
  \includegraphics[width=0.32\textwidth]{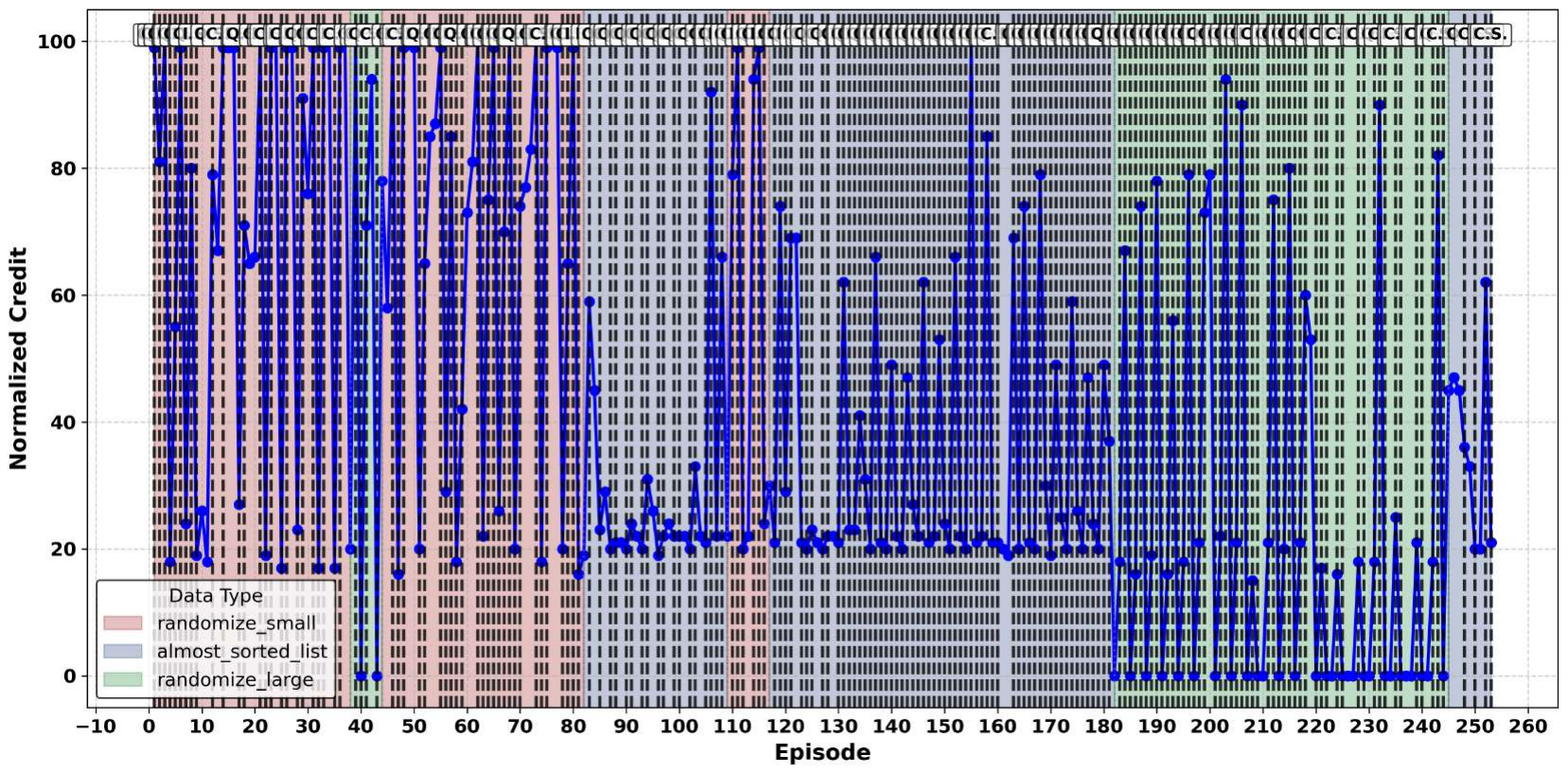}
}\hfill
\subfloat[$I_2$ (G) (Start Algo.: IS)]{
  \includegraphics[width=0.32\textwidth]{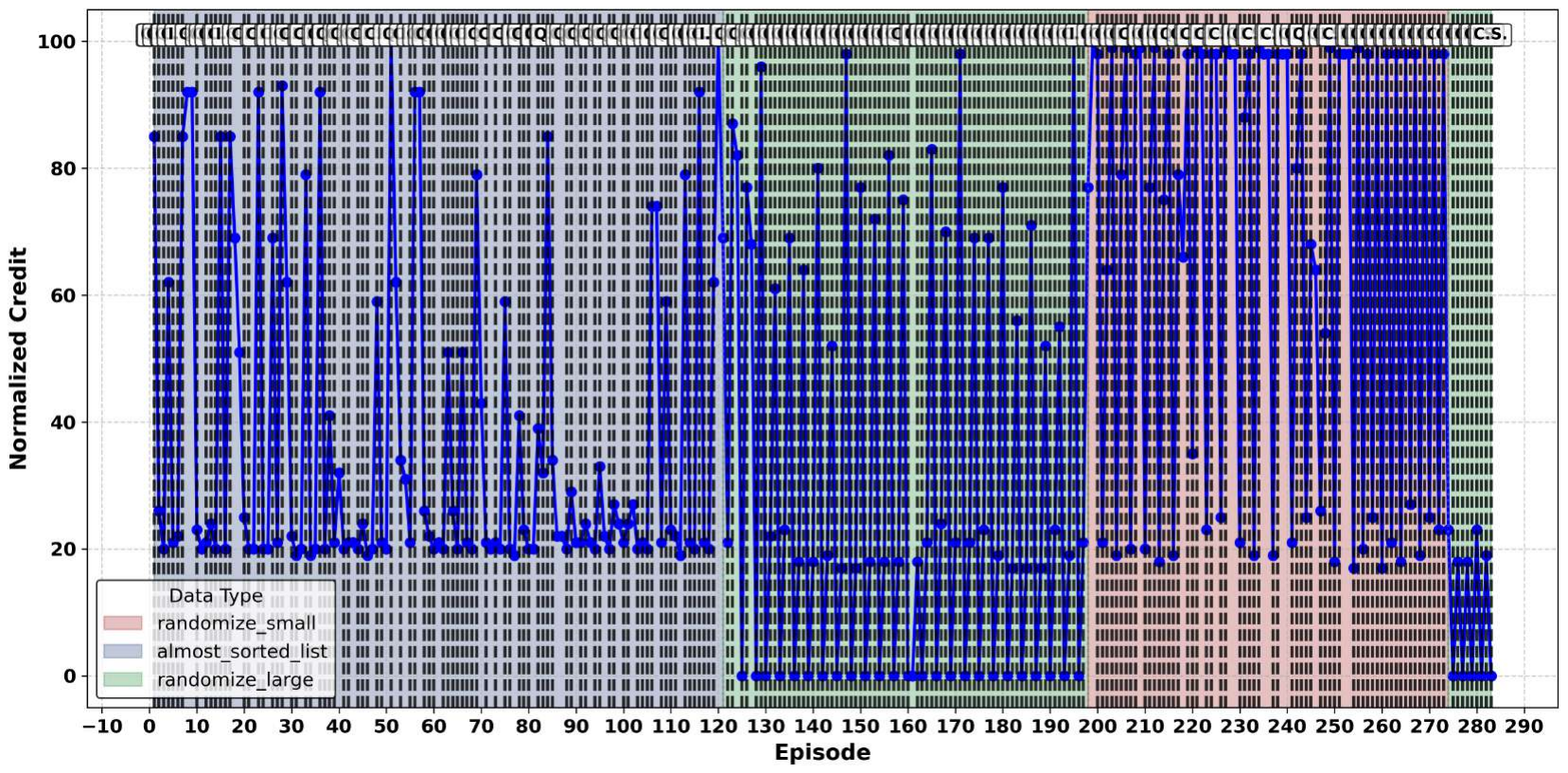}
}\hfill
\subfloat[$I_3$ (Start Algo.: QS)]{
  \includegraphics[width=0.32\textwidth]{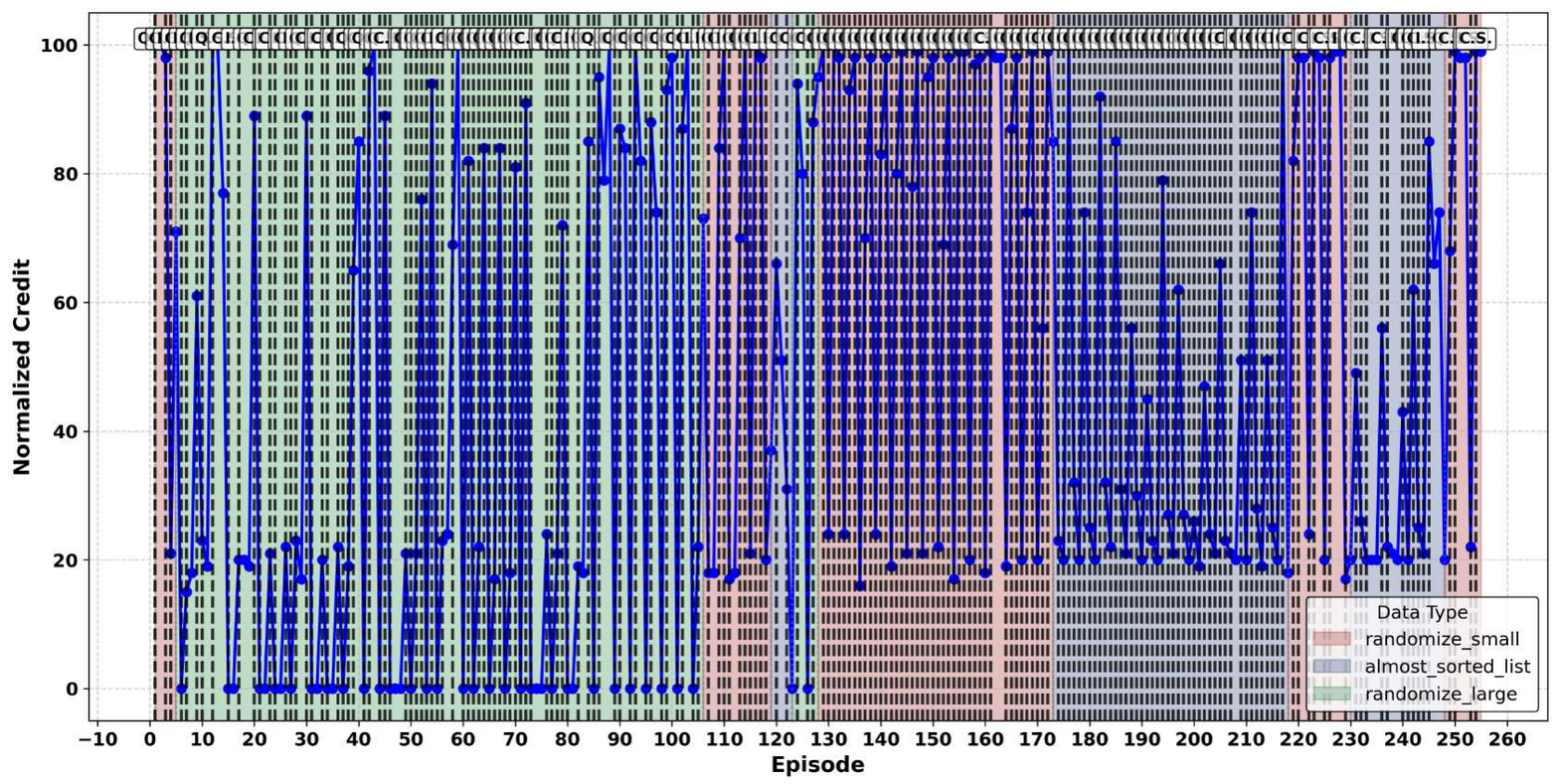}
}
}

\caption{Variations in normalized credits versus episodes in islands for the sorting scenario without and with the G-Island. Vertical dashed lines indicate switches to a new algorithm, while coloured regions indicate input data types. (G) denotes the G-Island.}

\end{figure}

\begin{figure}[t]
\centering

\textbf{With Yielory and No G-Island}\\[4pt]

\makebox[\textwidth][c]{%
\subfloat[$I_1$ (Start Algo.: CS)]{
  \includegraphics[width=0.32\textwidth]{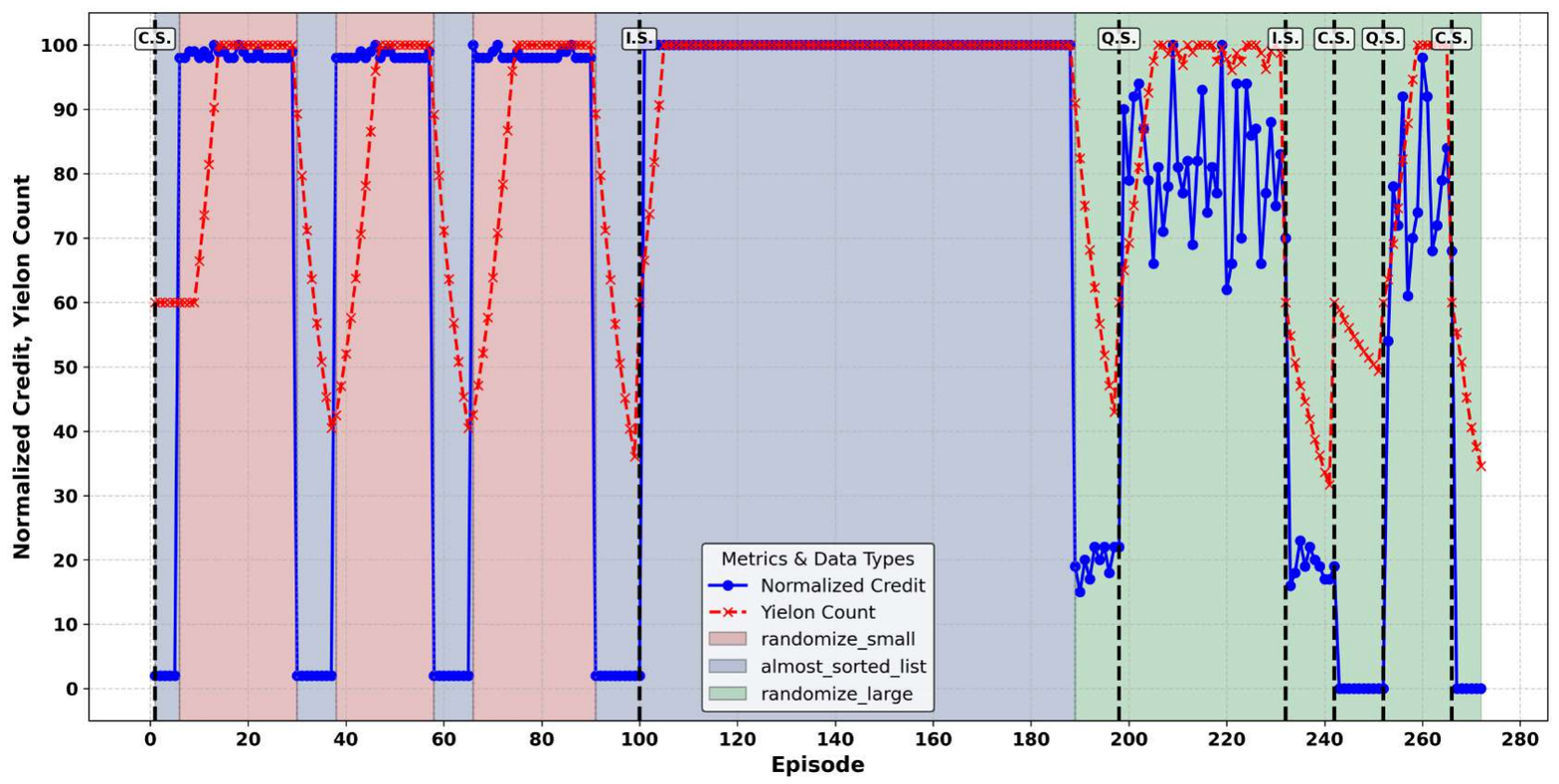}
}\hfill
\subfloat[$I_2$ (Start Algo.: IS)]{
  \includegraphics[width=0.32\textwidth]{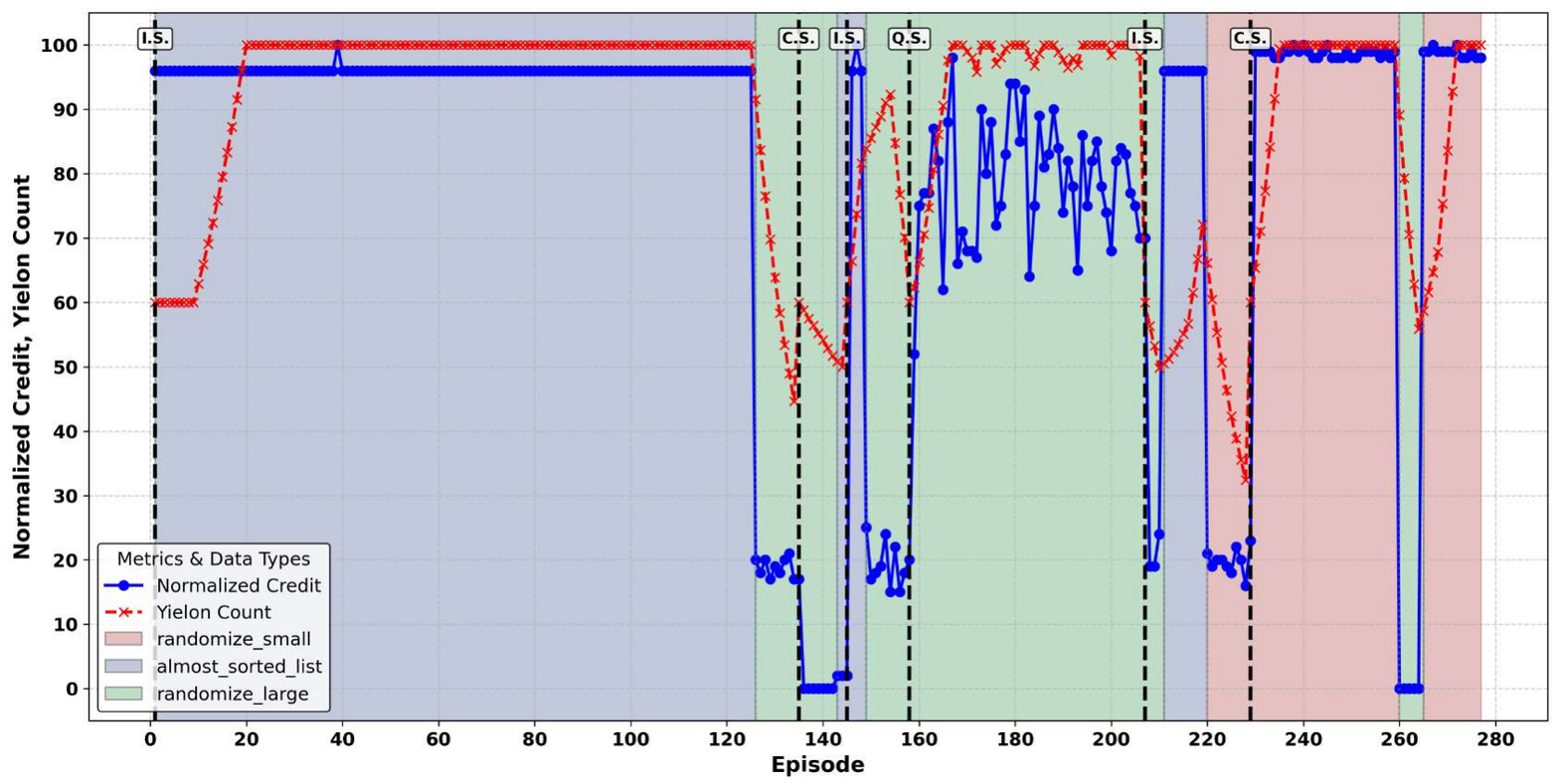}
}\hfill
\subfloat[$I_3$ (Start Algo.: QS)]{
  \includegraphics[width=0.32\textwidth]{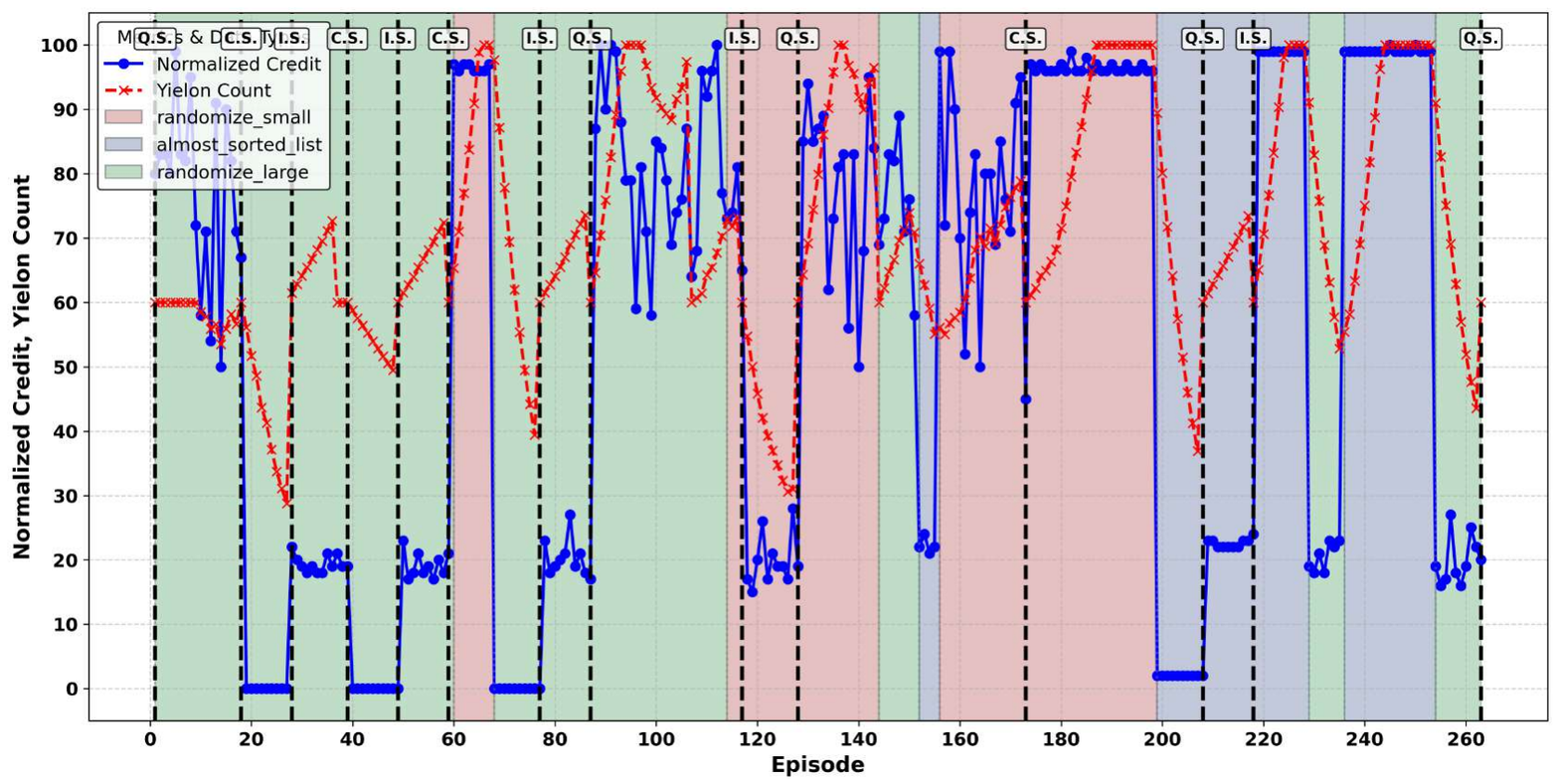}
}
}

\vspace{1em}

\textbf{With Yielory and G-Island}\\[4pt]

\makebox[\textwidth][c]{%
\subfloat[$I_1$ (Start Algo.: CS)]{
  \includegraphics[width=0.32\textwidth]{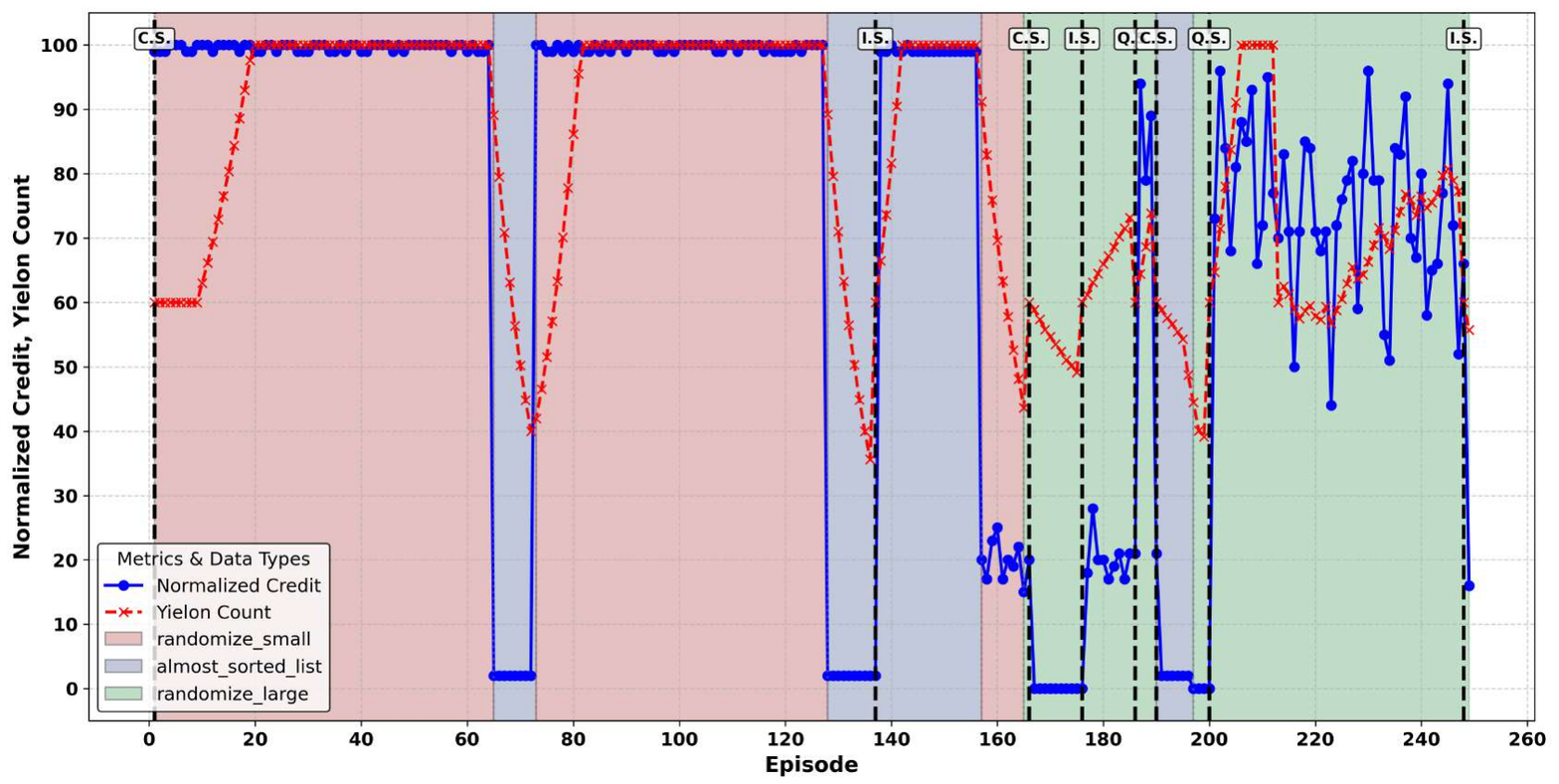}
}\hfill
\subfloat[$I_2$ (G) (Start Algo.: QS)]{
  \includegraphics[width=0.32\textwidth]{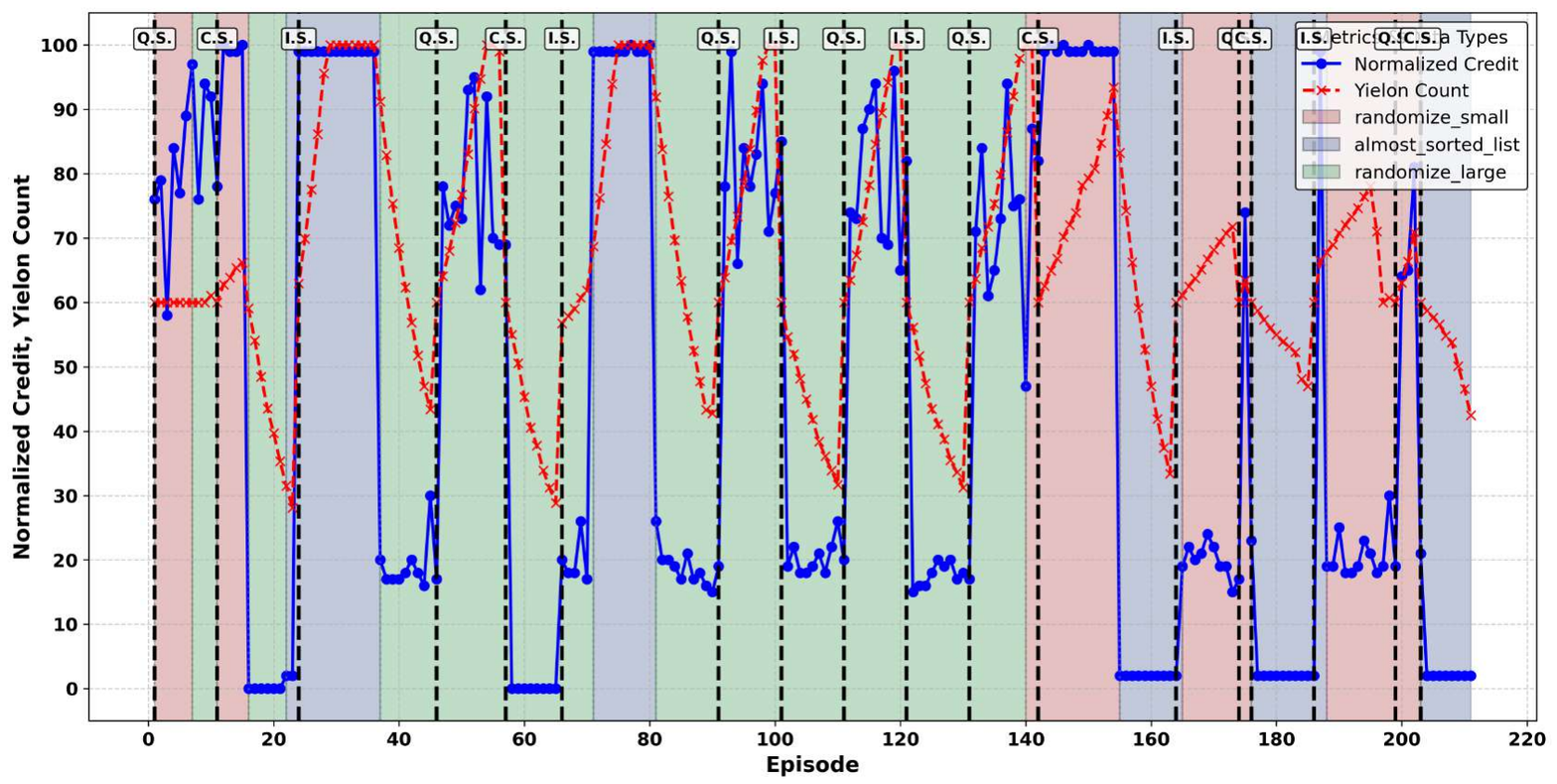}
}\hfill
\subfloat[$I_3$ (Start Algo.: IS)]{
  \includegraphics[width=0.32\textwidth]{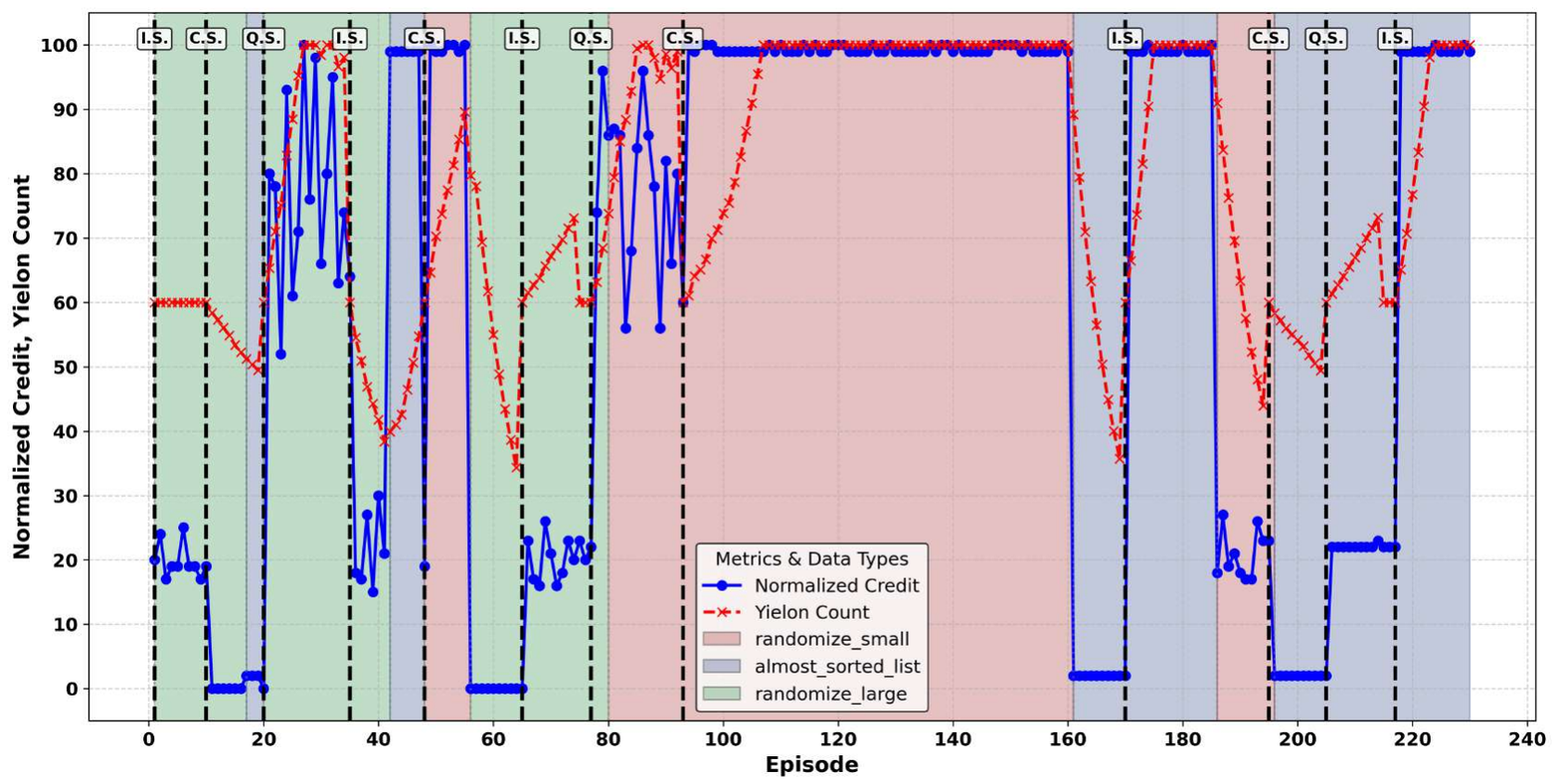}
}
}

\caption{Variations in normalized credits and Yielon counts versus episodes in islands for the sorting scenario without and with the G-Island. Vertical dashed lines indicate switches to a new algorithm, while coloured regions indicate input data types. (G) denotes the G-Island.}

\end{figure}

\begin{table}[ht]
\caption{Comparison of Normalized Credits \& Switches across Islands}
\scriptsize
\centering
\begin{tabular}{|c|l|c|c|c|c|c|c|}
\hline
\textbf{S.No.} &
\multirow{2}{*}{\hspace{0.5cm} Experiments Using Sorting}
& \multicolumn{2}{c|}{$I_1$}
& \multicolumn{2}{c|}{$I_2$ (G)}
& \multicolumn{2}{c|}{$I_3$} \\ \cline{3-8}

& & Credits & Switches
& Credits & Switches
& Credits & Switches \\ \hline

1 & No Yielory and No G-Island   &10384 &0 &3058 &1 &8491 &1 \\ \hline
2 & No Yielory and With G-Island &9569 &160 &7241 &163 &7411 &163 \\ \hline
3 & With Yielory and No G-Island &16019 &2 &15987 &3 &11102 &14 \\ \hline
4 & With Yielory and G-Island    &14469 &6 &9672 &17 &12663 &11 \\ \hline

\end{tabular}
\label{tab:client_comp}
\end{table}

Table 1 provides a comparison of normalized credits accumulated at the islands together with the number of switches they made during the runs. From the first two entries, (greedy approaches - with and without the G-Island) it can be observed that the introduction of the G-Island (which causes a huge amount of switching) does not contribute much to increase in Normalized credits. On the contrary, the introduction of the Yielory and the G-Island (3rd and 4th entries) show a substantial increase in the number of Normalized credits accumulated. It may also be noted that the introduction of the G-Island increased the number of switches at a marginal cost of Normalized credits in one of the islands ($I_1$), while for another it improved the same. The take-home is that the Yielory and the G-Island together, drastically decreased the number of switches when compared to those in the greedy approach while also ensuring a balance between exploitation and exploration. It is therefore clear that optimizing the ratio of the number of G-Islands to that of other islands can offer greater insights and improve performance.

\subsection{Obstacle Avoidance} 
The graphs in Figs. 5a - 5d depict variations in normalized credits and Yielon count at each episode in the four islands. As can be seen in all the graphs except to an extent in Fig. 5b, algorithm-switches allow the learning to absorb variations in input in spite of sudden changes in the normalized credits. While using learning algorithms that make use of models or policies, as in this case, it is essential that they be given sufficient opportunities to sample and learn from the input before deeming them to be non-performing. As expected, the three islands - 1, 3, and 4 - switch comparatively lesser number of times (Figs. 5a, 5c, and 5d) when compared to the more active G-Island (Fig. 5b). Together, these four islands ensure that saturation does not occur too soon and thus avoid local optima. As a rudimentary ablation study exercise to understand the significance of the G-Island, we removed the same and in lieu, placed a normal island, making all the islands in the archipelago homogenous. It was found that in both the cases - Sorting and Obstacle avoidance - the number of extrinsic explorations reduced, intuitively suggesting that the search seemed to have been curtailed to the algorithms in the respective local repertoires.

\begin{figure}[ht]
\centering  

\subfloat[$I_1$ (Start Algo.: Q-Learning)]{
  \includegraphics[height=2.9cm,width=0.45\textwidth]{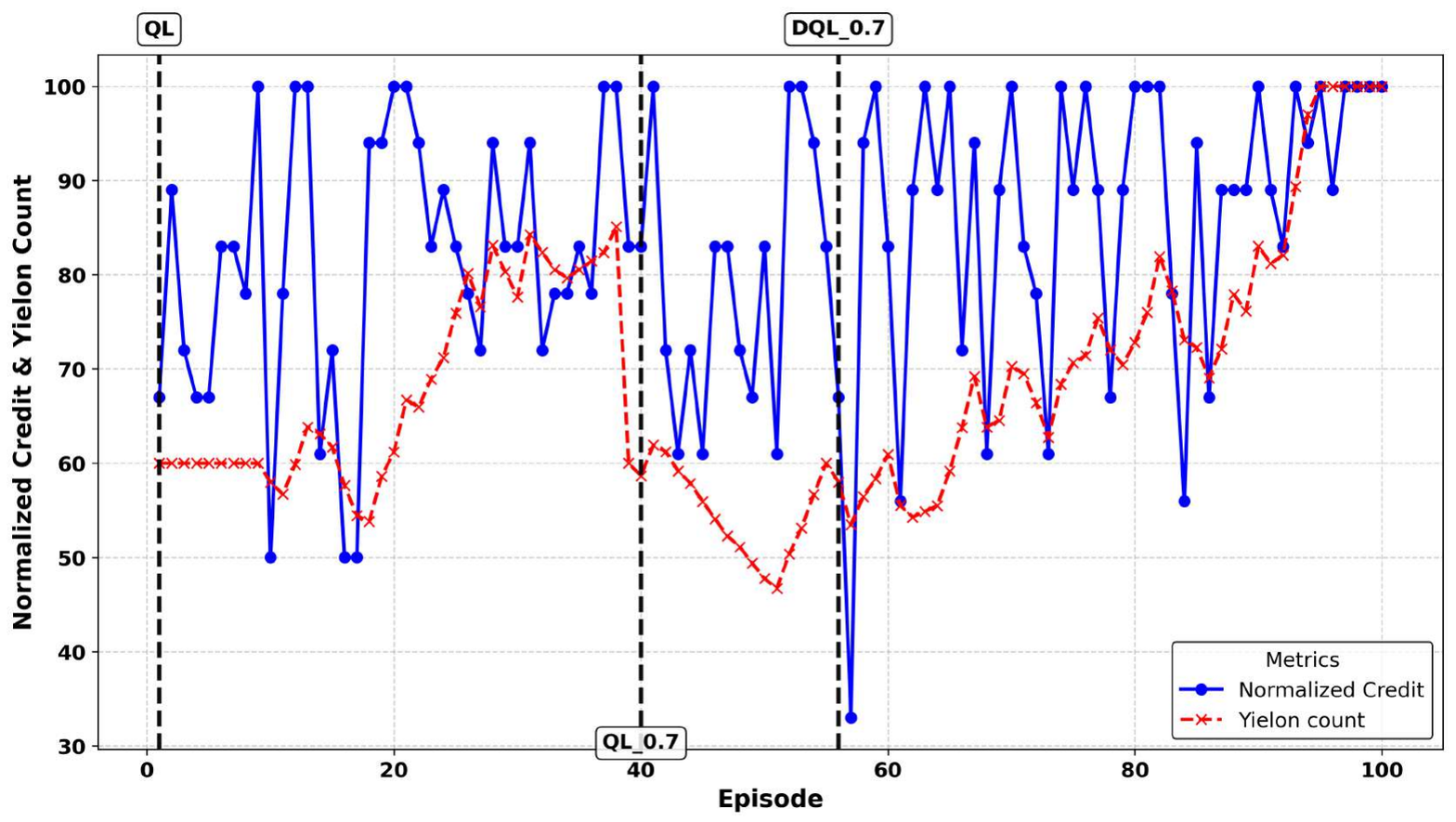}
}
\hfill
\subfloat[$I_2$ (G) (Start Algo.: SARSA)]{
  \includegraphics[height=2.9cm,width=0.45\textwidth]{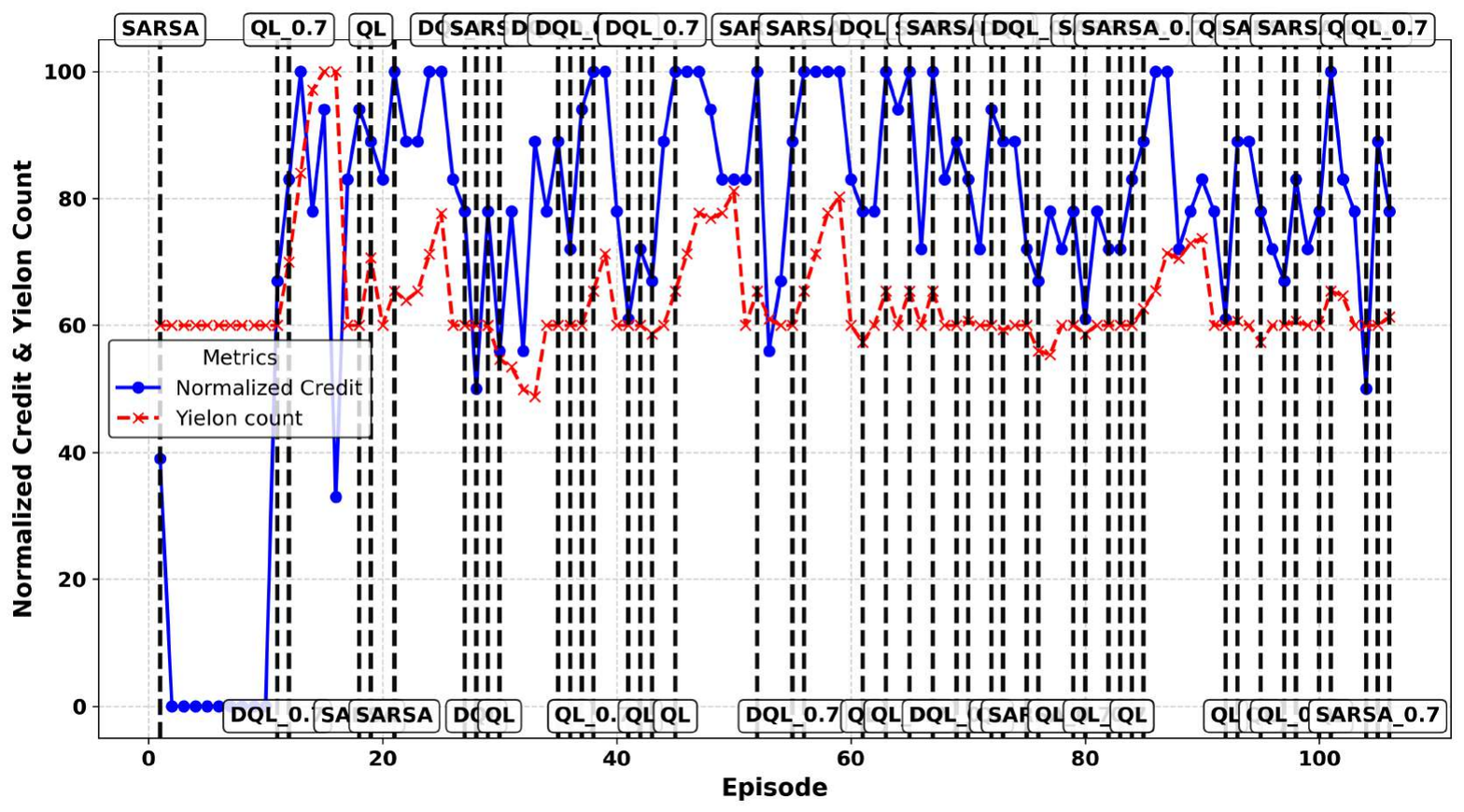}
}

\medskip

\subfloat[$I_3$ (Start Algo.: Double Q-Learning)]{
  \includegraphics[height=2.9cm,width=0.45\textwidth]{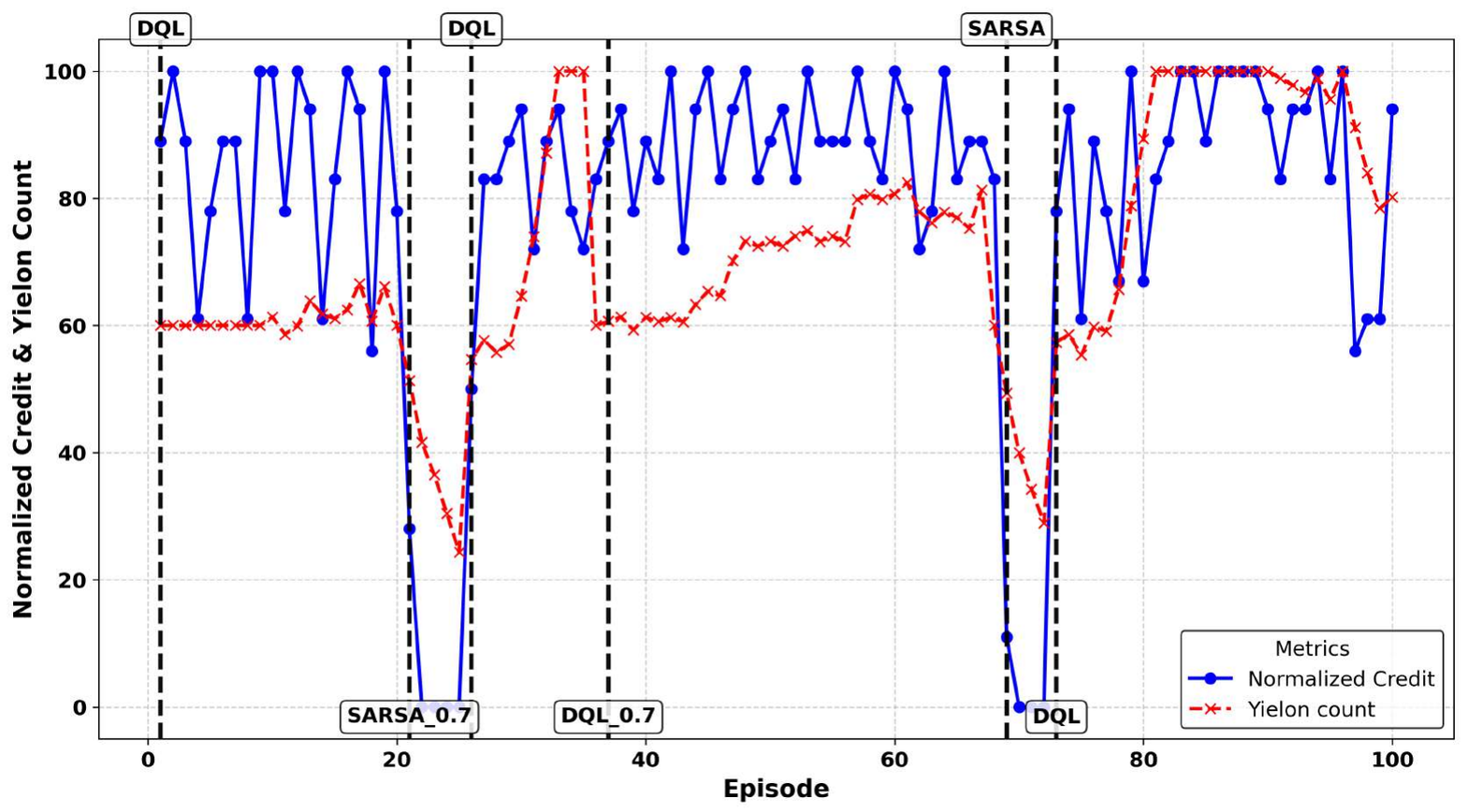}
}
\hfill
\subfloat[$I_4$ (Start Algo.: Q-Learning, $\alpha=0.7$)]{
  \includegraphics[height=2.9cm,width=0.45\textwidth]{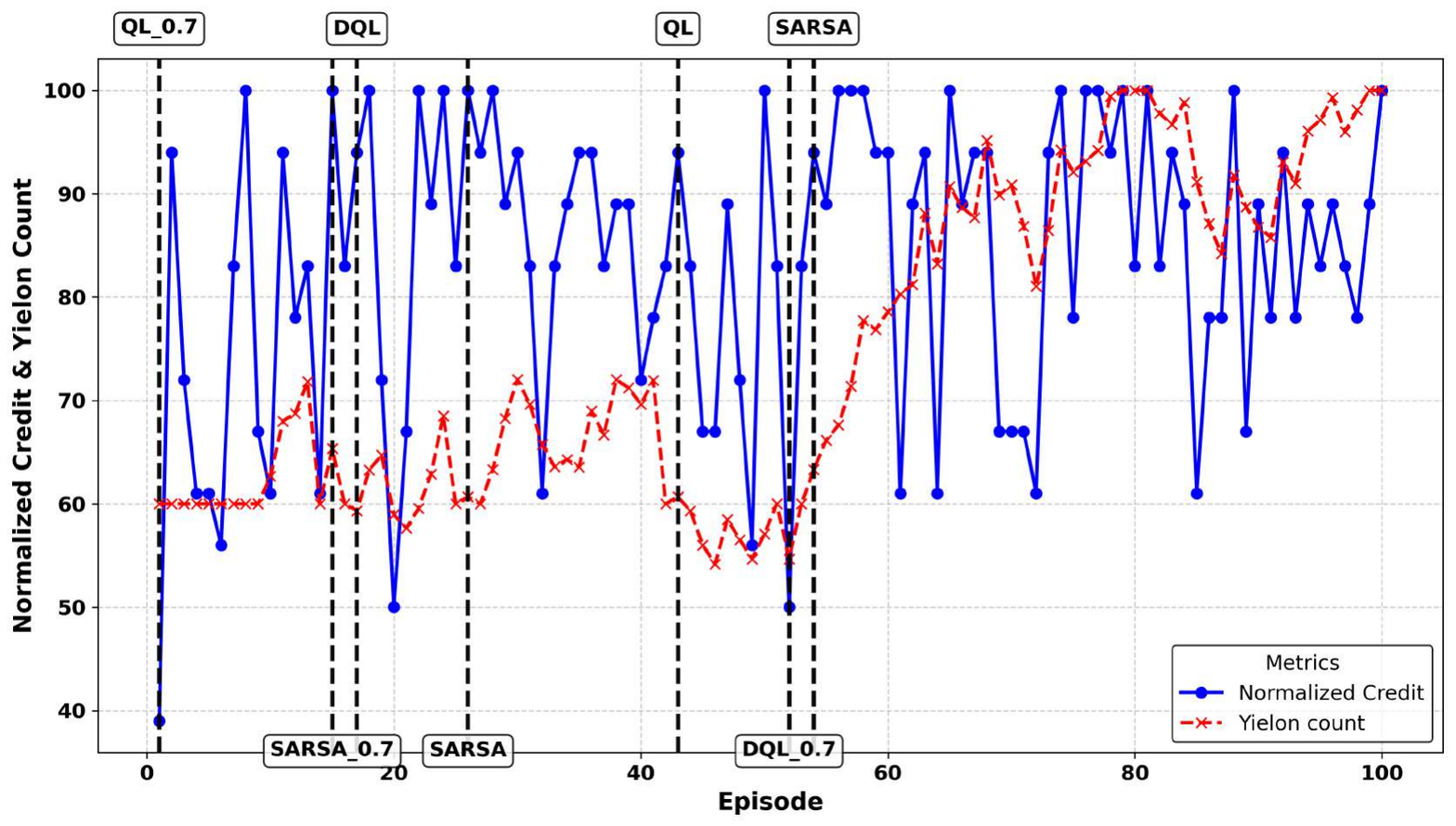}
}

\caption{Graphs depicting variations in Normalized Credits and Yielon Counts over Episodes for the Obstacle Avoidance Robotics Scenario in the four islands. (G) indicates the G-Island}
\end{figure}

\section{Conclusions and Future Work}
This paper describes an adaptive algorithm selection technique inspired by both a natural process and RL, which involves agents working in both isolation as well as in co-operation. Agents within islands containing a repertoire of algorithms, concurrently search for the best possible algorithm for task instances streaming in, thereby exploiting parallelism and hastening convergence. While the agents exploit the current algorithm as long as there are sufficient Yielons in the Yielory, they switch to exploring algorithms within their internal repertoires (\text{intrinsic exploration}) when the Yielon counts go below a threshold. Over time, when the Yielon count stabilizes, the agent is made to feel the occurrence of a local optima, forcing it to explore algorithms across islands and import the one that is the best, thereby exhibiting \text{extrinsic exploration}.The sharing of algorithms across islands in such extrinsic explorations, accelerated by a hyperactive G-Island, hints at genetic exchanges, which in turn mitigates the occurrence of evolutionary stagnation and premature convergence to suboptimal solutions as in the Island syndrome \cite{BAECKENS2020R338}. The content in the Yielory thus, controls exploitation and exploration during the overall search for the best algorithm and facilitates life-long learning. Results obtained from experiments carried out in the closed world (Sorting) and in an open world (Robotics) clearly demonstrate the efficacy of the proposed technique.
With more computing power, an island could be populated with multiple cooperative agents that feed on algorithms in the local repertoire, thereby speeding up convergence. 
Flushing ineffective or less performing algorithms across the repertoires by maintaining local Halls-of-Fame, as described in \cite{10.1145/3377929.3398149}, will also aid in reducing the search space.
Further, the proposed system could be used for life-long learning if new algorithms for either same or heterogeneous tasks were to be added to the repertoires on-the-fly \cite{10.1145/2783449.2783469}. The agents in the islands would then have the ability to find better algorithms that can tackle multiple tasks. In addition to algorithms being exchanged across islands during extrinsic explorations, hyper-parameters, too, could be made to cross-over or mutate, as in PBT \cite{jaderberg2017populationbasedtrainingneural}, to ameliorate the performance of a currently active algorithm.
In the future, we intend to increase the number of islands and incorporate Federated Learning (FL) using mobile agents, as in \cite{jp}, that can constantly migrate across the islands in search of better-performing learning models, aggregate them, and replant their modified versions in the repertoires within the islands. Such agents will make the CIA redundant and transform the system into a decentralized one where all three forms of learning - Island GA models, PBT, and FL - work in unison. When used in Multi-Mobile-Robot-Systems, such a technique will also empower robots to emulate the mobile islands whose network topography could constantly change due to their mobility in the physical environment.

\end{document}